\documentclass[prl,nofootinbib,twocolumn,floatfix,showpacs]{revtex4}
\usepackage{graphicx}
\usepackage{amsmath}
\usepackage{amssymb}

\begin{document}
\title{Macroscopic loop formation in circular DNA denaturation}
\date{\today }
\author{Alkan Kabak\c{c}{\i}o\u{g}lu$^{1}$, Amir Bar$^{2,3}$, David Mukamel$^{2}$\emph{}\\
\emph{$^{1}$Department of Physics, Ko\c c University, Sar\i yer 34450 \. Istanbul, Turkey}\\
\emph{$^{2}$Department of Physics of Complex Systems and }\\
\emph{$^{3}$ Department of Computer Science and Applied Mathematics,
The Weizmann Institute of Science, Rehovot 76100, Israel}}

\begin{abstract}
  The statistical mechanics of DNA denaturation under fixed linking
  number is qualitatively different from that of the unconstrained
  DNA. Quantitatively different melting scenarios are reached from two
  alternative assumptions, namely, that the denatured loops are formed
  in expense of 1) overtwist, 2) supercoils. Recent work has shown
  that the supercoiling mechanism results in a BEC-like picture where
  a macroscopic loop appears at $T_c$ and grows steadily with
  temperature, while the nature of the denatured phase for the
  overtwisting case has not been studied.  By extending an earlier
  result, we show here that a macroscopic loop appears in the
  overtwisting scenario as well. We calculate its size as a function
  of temperature and show that the fraction of the total sum of
  microscopic loops decreases above $T_c$, with a cusp at the critical
  point.

\end{abstract}


\maketitle

The melting or denaturation of DNA refers to the separation of the two
complementary DNA strands, a process which can be induced either
thermally~\cite{wartell1985thermal} or
mechanically~\cite{rouzina2001force,cocco2001force,marenduzzo2001phase}. The
associated phase transition is well understood by means of theoretical
models, e.g., by Poland and Scheraga~\cite{PS1966}, Peyrard and
Bishop~\cite{peyrard1989statistical}, and their various
extensions~\cite{hanke2008denaturation,dauxois1993dynamics,blake1999statistical}. Thermal
melting of DNA forms the basis of the PCR technique, while statistical
and dynamical properties of denatured loops may turn out to be relevant
for understanding DNA-protein
interactions~\cite{altan2003bubble,forties2009flexibility} and gene
expression initiation~\cite{kalosakas2004sequence,van2005can}.

Interestingly, it is a common practice to use plasmid (circular) DNAs
during PCR, since most bacteria come with circular DNA as a means of
protection against degradation. The resulting entanglement of the two
strands due to the natural twist of the DNA molecule imposes an
obvious obstacle for the denaturation process (as well as replication,
protein synthesis, etc.), which is overcome in nature by means of
special DNA manipulating
proteins~\cite{alberts2004essential}. Nevertheless, the thermal
behavior of a circular DNA chain in absence of such helper proteins
proves to be a nontrivial problem and has been addressed
recently~\cite{RB2002,GOY2004,KOM09}.

The circular geometry entails the presence of a new topological
invariant in the system: the number of times two chains of the DNA
wind around each other, namely, the linking number (LK).  The
thermodynamics of the system should therefore be investigated within
the corresponding, restricted phase space. This framework is also
relevant to single-molecule experiments on DNA in which the chain ends
are rotationally constrained~\cite{bryant2003structural}. We here
discuss the implications of LK conservation on thermal melting
characteristics within the framework of the Poland-Scheraga (PS) model.

A consequence of fixing the linking number is that, the denatured
loops form in expense of (right-handed) torsional stress on
surrounding DNA duplex segments. As with any elastic ribbon with
finite bending and twist modulus, dsDNA responds to torsion by
``supercoiling'' (bending the backbone as in coiling telephone cords)
and/or by ``overtwisting'' (modifying the stacking
angle). Thermodynamics of a fixed-LK DNA chain whose bound segments
are unbendable but have finite twist rigidity was investigated by
Rudnick \& Bruinsma~\cite{RB2002}. The alternative extension of the PS
model considering the possibility of supercoil formation, but not
overtwisting, has also been discussed
recently~\cite{KOM09,bar2011supercoil}. It has been shown that the
transition is a kind of Bose Einstein condensation (BEC) where the
macroscopic loop formed above the melting temperature plays the role
of the condensate.

We show below that a BEC-like transition takes place in the
``overtwisting'' scenario, too, which is the main contribution of this
article. We next compare this phenomenon with a similar observation we
made earlier for the supercoiling response and with the denaturation
of DNA with free ends. We conclude that, the birth of a nontrivial
macroscopic loop at the melting point is the defining characteristics
of the thermal denaturation of DNA under fixed linking number,
irrespective of how the molecule responds to torsional stress.

Ref.\cite{kafri2000dna} gives a detailed account of the melting
transition in the PS model. The partition function of the model can be
expressed in closed form and its singular behavior, with proper
treatment of the loop entropy, yields a first-order melting
transition. A similar analysis is given in
Ref.\cite{KOM09,bar2011supercoil} for the case of nonzero supercoil
density under the constraint that the total length of the denatured
loops is proportional to that of supercoils (mimicking LK
conservation). This system, unlike the PS model, displays a continuous
melting transition, accompanied by a loop ``condensate'' that appears
at $T_c$ and grows gradually with temperature.

Overtwisting, i.e., increasing the stacking angle between the
successive base pairs, is the alternative to supercoiling by which a
partially denatured circular DNA chain can accommodate the resulting
torsional stress on duplex regions. An extension of the PS model with
overtwist has been investigated earlier in \cite{RB2002}, which we
extend below. In particular, we show here that the melting transition
in this scenario, too, is accompanied by the formation a macroscopic
loop.

To this end, following \cite{RB2002}, let us consider an arbitrary
configuration of a DNA chain of a total length $L$ composed of
denatured segments with total length $L_l$ and bound segments with
$L_b=L-L_l$. We assume that the linking number expelled by the loops
is uniformly distributed along the chain and results in a uniform
increase $\delta\theta \propto L_l/L_b$ in the stacking angle measured
between the two ends of a unit DNA segment. Associated overtwist
energy per unit length is assumed harmonic and the total internal
energy can be written as
\begin{eqnarray}
\label{Hamiltonian}
{\cal H} &=& \kappa\, \frac{L_l^2}{L_b} + \epsilon_b L_b,
\end{eqnarray}
where $\kappa>0$ is a measure of overtwist stiffness in units of
energy and $\epsilon_b<0$ is the binding energy per unit length. The
canonical partition function for the DNA chain can then be expressed
as \cite{RB2002}
\begin{eqnarray}
\label{contour_int}
 Z^\kappa(L_b,L_l) &=& \oint \frac{dz_b dz_l}{(2\pi i)^2}\
 \frac{Q^{\kappa=0}(z_b,z_l)}{z_b^{L_b+1}z_l^{L_l+1}}
\  e^{-\beta\kappa\, \frac{L_l^2}{L_b}},
\end{eqnarray}
where the grand sum $Q^{\kappa=0} = [1/\omega z_b -1 - A\Phi_c(s
  z_l)]^{-1}$ follows from the usual PS model with the Boltzmann
weight $\omega$ for a unit bound segment and fugacities $\{z_b,z_l\}$
per unit length of bound and denatured DNA, respectively. The contour
integral has a simple pole for $z_b$ which by Cauchy formula yields
\begin{eqnarray}
\label{contour_eval}
Z^\kappa(L_b,L_l) &=& \frac{1}{2\pi i} \oint \frac{dz_l}{z_l^{L_l+1}}\,
e^{-\beta\kappa\, \frac{L_l^2}{L_b}} \omega^{L_b}\nonumber \\
&&\times\bigg[1+A\Phi_c(sz_l)\bigg]^{L_b-1} \equiv \oint {dz_l} e^{-L F(z_l,m_l)}, \nonumber \\
\mbox{with}\ \ \ \ \ \ \ \ && \nonumber \\
F(z_l,m_l)\ \  &=& -(1-m_l)\log\bigg[\omega [1 + A\Phi_c(sz_l)]\bigg]\nonumber \\
&& + m_l \log z_l + \beta\kappa \frac{m_l^2}{1-m_l} + O(L^{-1})\ ,
\label{eq_F}
\end{eqnarray}
where $m_{l,b}=L_{l,b}/L$ and $m_l + m_b = 1$. In the thermodynamic
limit, the partition function can be evaluated using the saddle-point
condition $\partial F = 0$. Therefore, $F(z_l,m_l)$ serves as a free
energy functional for the DNA chain.  Minimization yields a continuous
phase transition for $c>2$ governed by the singularity of the polylog
function at $sz=1$~\cite{RB2002}. It is straightforward to show the critical
temperature $T_c$ shifts linearly with the ratio of the overtwist
penalty $\kappa$ and the binding energy $\epsilon_b =
-k_BT\log\omega$:
\begin{eqnarray}
T_c &=& T_c^{PS}\,\bigg[1 - \frac{\kappa}{\epsilon_b}\bigg(\frac{1}{(1-m_l^c)^2}-1\bigg)\bigg]\ . \label{eq_Tc}
\end{eqnarray}
Here $T_c^{PS}$ is the critical temperature of the original PS model
and $m_l^c$ is the critical loop fraction which we find to be
independent of the twist stiffness (therefore equal to the
corresponding value in the PS model):
\begin{eqnarray}
\label{eq_ml}
m_l^c &=& \frac{A\zeta_c}{1+A(\zeta_c+\zeta_{c-1})}\ .
\end{eqnarray}
Here $\zeta_c = \Phi_c(1)$ is the Riemann zeta function.

Our goal is to investigate the existence of a macroscopic loop for
$T>T_c$ in the above picture. Let us assume that such a loop exists
with size $L_0 \equiv m_0L$ and calculate $m_0$. As for a Bose gas
below the condensation temperature, the estimated amount of denatured
DNA, when calculated as a sum over microscopic loops, is now short of
the actual value by $L_0$. Therefore we set $L_l = L_l^{micro}+L_0$
and substitute in Eq.\ref{contour_int} \[Z^\kappa(L_l,L_b) \rightarrow
Z^\kappa(L_l-L_0,L_b)\,\frac{s^{L_0}}{L_0^c}\] which, following the
same steps, now yields an additive macroscopic loop correction to
Eq.\ref{eq_F}:
\begin{eqnarray}
F(z_l,m_l,m_0) &=& F(z_l,m_l)-m_0\log(sz_l) \label{eq_newF}\ .
\end{eqnarray}
Note that $m_l$ is the total loop density, including microscopic {\it
  and} macroscopic contributions.

For $T<T_c$ ($z_l<1/s$), the free energy is minimized {\it wrt} $m_0$
at the extremal value $m_0=0$. Therefore no macroscopic loop exists
below $T_c$. The total fraction of denatured bases (all due to
microscopic loops) can be calculated by setting $\partial_{m_l}F =
\partial_{z_l}F = 0$. It increases with temperature slower than in the
PS model, due to the additional (overtwist) energy penalty of
denaturation. The loop-size distribution essentially decays
exponentially with a power-law correction as in the PS model.

\begin{figure*}
  \centering
  \includegraphics[width=0.7\linewidth]{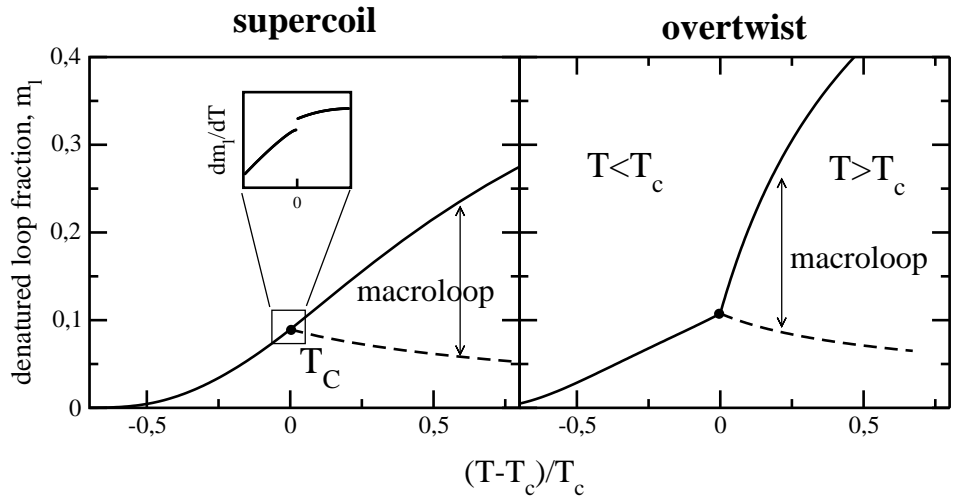}
  \caption{\label{Fig1} The fraction of the denatured DNA (solid) and
    the contribution from the microscopic loops (dashed) for the
    supercoiling (left) and overtwisting (right) scenarios, as a
    function of temperature. The parameters $s=5$, $\epsilon_b=3$,
    $c=3.5$, $A=0.1$ are the same for both figures, while the
    stiffness parameters for overtwisting ($\kappa=1.0$) and
    supercoiling (see Ref.\cite{bar2011supercoil}) are comparable.}
\end{figure*}

In the high-temperature phase where the loop fugacity is fixed at its
upper bound $z_l=1/s$, the above picture is no longer valid. Setting now
$\partial F(z_l,m_l,m_0)=0$ yields a unique solution with $m_0\ne 0$
(the extremal value $m_0=0$ does not yield a minimum.)
Therefore, a finite fraction of the base pairs is located in a
macroscopic loop. Setting $\partial F/\partial m_0=0$ yields
$z_l=1/s$, as expected in the high-temperature phase. Accordingly, the
probability distribution function for the {\it microscopic} loop sizes
$p(l) \sim l^{-c}$ is now scale invariant. This picture is in contrast
with the PS model, since no microscopic loop survives for
$T>T_c$. An analytical expression for the mass fraction in microloops
and the macroloop can be obtained from the remaining two minimization
conditions {\it wrt} $m_l$ and $z_l$:
\begin{eqnarray}
0 &=& \frac{\partial F}{\partial m_l}\ \ \Rightarrow\ \ \frac{\omega}{s}\,(1+A\zeta_c) = e^{-\beta\kappa[-1 + 1/(1-m_l)^2]} \label{eq_dFdml}\\
0 &=& \frac{\partial F}{\partial z_l}\ \ \Rightarrow\ \ m_l = \frac{m_0+R}{1+R}, \label{eq_dFdzl}
\end{eqnarray}
where $R\equiv \frac{A\zeta_{c-1}}{1+A\zeta_c}$ is a
temperature-independent constant. The transition is continuous, since
setting $m_0=0$ ($T\rightarrow T_c^+$) recovers the critical value of
$m_l$ in Eq.(\ref{eq_ml}) found from the low-temperature limit $T\to
T_c^-$. Solving for $m_0$ in Eqs.(\ref{eq_dFdml}\&\ref{eq_dFdzl}),
then substituting $\ln\omega = -\epsilon_b/k_BT$ and using
Eq.(\ref{eq_Tc}) one finds that the macroscopic loop fraction has the
exact form
\begin{eqnarray}
m_0 &=& 1- \frac{1}{\sqrt{1+B(T-T_c)}},
\end{eqnarray}
where $B = \frac{-\epsilon_b}{\kappa T_c}(1+R)^{-2}$. $m_0$ grows
linearly with temperature in the vicinity of $T_c$ and approaches
unity only in the limit $T\to\infty$. The total fraction of
microscopic loops decreases with temperature while their distibution
remains a power law. In Fig.\ref{Fig1} we present a comparison of the
``overtwist'' and the ``supercoil'' dominated scenarios in terms of
macroscopic and microscopic loop fractions, for a generic set of
parameters at which both display a second-order melting transition.
Note that, the discontinuity in $dm_l/dT$ at $T=T_c$ (inset) in the
supercoiling model is hardly visible. However, the cusp in the
microscopic loop fraction is clear and exists even for $c<3$ where the
discontinuity at $T_c$ shifts to higher derivatives of the free energy
(see Ref.\cite{bar2011supercoil} for more on PS model with
supercoiling). The cusp in $m_l$ is more prominant in the
``overtwist'' picture with a comparable set of parameters.

\begin{figure}[b!]
\vspace*{10pt}
  \begin{center}
      \includegraphics[width=7cm]{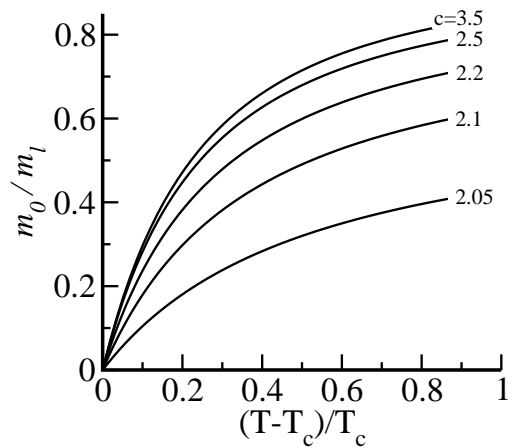}
  \end{center}
  \caption{\label{Fig2} The weight of the macroloop among the total
    amount of denatured bases is shown as a function of temperature,
    for various values of $c$. The uppermost curve for $c=3.5$ is
    indistinguishable from those obtained for higher values of $c$.}
\end{figure}

The presence of the macroloop above $T_c$ does not depend on the
precise value of $c$ (as long as there's a melting transition), while
its size relative to the total amount of denatured DNA, $m_0/m_l$,
does (Fig.\ref{Fig2}). For $c\le 2$, the macroscopic loop vanishes
together with the transition itself. Otherwise, the relative size of
the macroloop grows faster with temperature as $c$ gets larger. This
trend as a function of $c$ converges to a limiting curve which is
indistinguishable from that obtained for $c=3.5$ and shown in
Fig.\ref{Fig2}.

Generalization of our results to $\sigma\equiv (L_s-L_l)/L \neq 0$ is
straightforward after substituting $\beta\kappa(m_l+\sigma)^2/(1-m_l)$
for the twist penalty in Eq.(\ref{eq_F}). The melting picture for
nonzero $\sigma$ remains qualitatively unaltered, except when $\sigma
\le -m_l^c$, where the critical point itself disappears and a
macroscopic loop exists at all temperatures.

\begin{figure}[h!]
  \begin{center}
      \includegraphics[width=8cm]{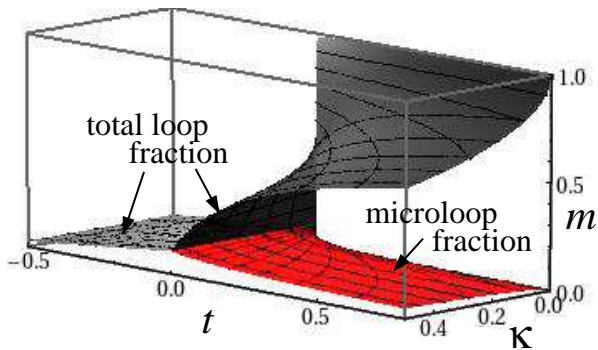}
  \end{center}
  \caption{\label{Fig3} (Color) The loop fraction $m$ of the DNA as a
    function of the reduced temperature $t=(T-T_c)/T_c$ and the
    overtwist penalty $\kappa$. The red surface shows the contribution
    of the microscopic loops to the total denatured DNA. The size of
    the macroscopic loop is the vertical distance between the upper
    (gray online) and the lower (red online) surfaces for $t>0$. The
    back panel of the bounding box corresponds to the PS model. Note
    that, the reduced temperature $t$ does not correspond to a single
    temperature value, since $T_c$ is a function of $\kappa$.}
\end{figure}

Finally, noting that the limit $\kappa\to 0$ recovers the first-order
transition, it is interesting to see how the sharp denaturation in the
PS model smoothens out with the introduction of twist penalty. This
crossover is shown in Fig.\ref{Fig3} where the total loop fraction
given in Fig.\ref{Fig1} is extended into the $\kappa$-dimension. The
region between the two sheets that join at the critical line $t=0$ is
the macroscopic loop fraction which smoothly approaches unity as
$\kappa\to 0$.

To summarize, we reconsidered the melting thermodynamics of a DNA
chain with fixed linking number. We assumed that the denatured loops
appear by transferring LK to duplex regions through overtwisting. We
showed that, despite the different melting scenarios observed in
supercoiling and overtwisting pictures, a feature common to both is
the appearance of a macroscopic loop which grows monotonously with
temperature. While the total fraction of denatured pairs increases
with temperature, the DNA mass in the {\it microscopic} loops
decreases above $T_c$. This condensation phenomenon is analogous to
BEC, except it takes place at high temperature. Whether it is
dynamically accessible is an interesting question, since merging
microscopic loops towards a macroloop entails diffusing denaturation
bubbles across torsionally strained duplex regions. Investigations in
this direction, as well as towards a joint theoretical framework that
incorporates both overtwist and supercoiling are in progress.

This work has been supported by the Scientific and Technological
Research Council of Turkey (TUBITAK) through Grant TBAG-110T618, and
by the Israeli Science Foundation (ISF).

\bibliography{paper}
\bibliographystyle{unsrt}

\end{document}